\documentstyle[12pt]{article}
\begin{document}
\parskip=5pt plus 1pt minus 1pt

\begin{flushright}
{\bf INS-Rep-1162}\\
{\bf DPNU-96-50}
\end{flushright}
\begin{flushright}
{September, 1996}
\end{flushright}

\vspace{0.2cm}
\begin{center}
{\large\bf A String-inspired Ansatz for Quark Masses and Mixing}
\end{center}
\vspace{0.5cm}

\begin{center}
{\sc Tatsuo Kobayashi} \footnote{ Electronic 
address: Kobayast@ins.u-tokyo.ac.jp} \\
{\it Institute for Nuclear Study, University of Tokyo, Tanashi, 
Tokyo 188, Japan}
\end{center}

\begin{center}
{\sc Zhi-zhong Xing} \footnote{ Electronic 
address: Xing@eken.phys.nagoya-u.ac.jp} \\
{\it Department of Physics, Nagoya University, 
Chikusa-ku, Nagoya 464-01, Japan}
\end{center}
\vspace{2.cm}

\begin{abstract}
We propose a simple but realistic pattern of quark mass matrices
at the string scale, which can be derived from orbifold models 
of superstring theory with no use of gauge symmetries. This pattern 
is left-right symmetric and preserves the structural parallelism 
between up and down quark sectors. Its phenomenological consequences 
on flavor mixing and $CP$ violation are confronted with current
experiments at the weak scale by use of the renormalization-group
equations in the framework of the minimal supersymmetric standard
model. We find that good agreement is achievable without fine-tuning.
\end{abstract}

\newpage

\section{Introduction}

The origin of fermion masses is one of the most important problems 
in particle physics.
Higher dimension couplings like $HQq(\theta /M)^n$ can explain 
the hierarchical structure of fermion masses and 
flavor mixing angles \cite{mass,RRR,IG}. 
The key point is that 
these couplings may provide effective Yukawa couplings with 
suppression factors $(\langle \theta \rangle /M)^n$, 
when suitable fields such as $\theta$ develop vacuum expectation 
values (VEVs).
Underlying symmetries provide selection rules for higher dimension couplings 
as well as renormalizable couplings.
In this case, the structure of powers $n$ of suppression factors 
are determined in terms of some types of quantum numbers 
from underlying symmetries.

\vspace{0.3cm}

For example, gauge symmetries including the anomalous $U(1)$ symmetry 
have been used to 
constrain such powers in nonrenormalizable couplings \cite{mass,IG}.
In this case one needs different quantum numbers for families under gauge 
symmetries, which are broken by VEVs of scalar fields, in order to obtain 
a realistic hierarchy of fermion masses.
On the other hand, through such symmetry breaking we have $D$-term 
contributions to soft scalar masses, which are proportional to 
quantum numbers of broken gauge symmetries \cite{Dterm}.
Thus different quantum numbers for families naturally resolve the
degeneracy of sfermion masses.
However, current measurements of the flavor changing neutral current 
(FCNC) processes require the degeneracy of sfermion masses, at least, 
for the first and second families of squarks \cite{fcnc}.

\vspace{0.3cm}

In the fermion mass generation mechanisms reviewed above, 
one needs the symmetries which lead to the hierarchical fermion 
mass matrices while preserving the required degeneracy 
of soft scalar masses without fine-tuning. 
Gauge symmetries are not such symmetries.
Superstring theory is the only known candidate for 
a successful unified theory of all the interactions including gravity. 
In superstring theory, selection rules 
for higher dimension couplings are provided by symmetries of 
a six-dimensional comactified space as well as gauge symmetries. 
These ``stringy symmetries'' due to the six-dimensional compactified space 
have no direct relation with soft scalar masses, and the fermion mass 
generation due to these stringy symmetries preserves the 
degeneracy of sfermion masses.
That is the difference between stringy symmetries and gauge symmetries.
In Ref. \cite{NR2}, 
selection rules for nonrenormalizable couplings are discussed 
explicitly within the framework 
of orbifold models \cite{Orbi}. 
These selection rules can lead to realistic quark mass matrices.
Actually one of simple but instructive mass matrices 
has been derived from a variety of orbifold 
models \cite{stringm} \footnote{Note that some different attempts 
have been made to derive the fermion mass matrices from other types 
of four-dimensional string models. See, e.g., Ref. \cite{stringm2}.}.
The pattern is left-right symmetric, and a parallel structure 
exists between up and down quark sectors.

\vspace{0.3cm}

It seems natural that left-right symmetric mass matrices have the
parallel structure for up and down quark sectors in the above-mentioned
fermion mass generation mechanism. The reason is quite straightforward:
if quark mass matrices are left-right symmetric, they are dominated 
by quantum numbers of the quark doublets $Q_i$ and the Higgs fields.
Of course, the up-down symmetric quark mass 
matrices can also be obtained through other approaches, e.g., 
by using an extra $U(1)$ symmetry \cite{IG}.

\vspace{0.3cm}

In this work we want to study a simple and realistic pattern of quark mass 
matrices at the string scale $M_{\rm st}$, which can be derived from 
orbifold models of superstring theory. 
This type of quark mass matrices is left-right symmetric and has the 
structural parallelism between up and down quark sectors.
By use of the renormalization-group equations in the framework of the
minimal supersymmetric standard model (MSSM) below $M_{\rm st}$, we confront the
consequences of our string-inspired mass pattern with current experimental
data at the weak scale $M_Z$. The flavor mixing matrix elements 
$|V_{us}|$, $|V_{cd}|$, $|V_{ub}/V_{cb}|$ and $|V_{td}/V_{ts}|$ can
be scale-independently obtained in terms of the quark mass ratios
$m_u/m_c$ and $m_d/m_s$ (as well as a $CP$-violating phase shift) in
leading order approximations. So are three inner angles of the unitarity
triangle $V^*_{ub}V_{ud} + V^*_{cb}V_{cd} + V^*_{tb}V_{td} =0$.
For reasonable values of $\tan\beta_{\rm susy}$ (the ratio of Higgs vacuum 
expectation values in the MSSM), we find that the renormalized 
$|V_{cb}|$ and $|V_{ts}|$ at $M_Z$ can fit current data very well.

\section{Quark mass matrices in superstring theory}

In general, the underlying theory like supergravity or superstring theory 
has nonrenormalizable couplings as  
\begin{equation}
h_{{\rm u}ij}H_2Q_iu_j(\theta_{\rm u}/M_2)^{n_{ij}} \; , ~~~~~~~~
h_{{\rm d}ij}H_1Q_id_j(\theta_{\rm d}/M_1)^{n'_{ij}} \; ,
\end{equation}
where $u_j$ ($d_j$) denotes the up type (down type) of $SU(2)$ singlet 
quark fields, and $H_{2,1}$ are 
the Higgs fields for the up and down sectors.
Here $h_{{\rm u}ij}$ and $h_{{\rm d}ij}$ denote the coupling strengths, 
which can be calculated within the framework of superstring theory.
Their magnitudes are of $O(1)$ in most cases.
When the fields $\theta_{\rm u,d}$ develop VEVs, these couplings become 
Yukawa couplings with suppression factors 
$\varepsilon_{\rm u}=(\langle\theta_{\rm u}\rangle/M_{2})^{n_{ij}}$ and 
$\varepsilon_{\rm d}=(\langle\theta_{\rm d}\rangle/M_{1})^{n'_{ij}}$. 
These can lead to a hierarchical structure in the fermion mass matrices.
Here we restrict ourselves to the case where up and down mass 
matrices have the parallel structure, i.e., $n_{ij}=n'_{ij}$.

\vspace{0.3cm}

The structure of the powers $n_{ij}$ depends on underlying symmetries.
If gauge symmetry breaking is dominant for the fermion mass generation, 
we obtain nondegenerate soft sfermion masses due to $D$-term 
contribution of broken symmetries in the case that fine-tuning is absent.
For example, nondegenerate soft scalar masses are in general 
derived from supersymmetry (SUSY) breaking within the superstring framework, 
except the dilaton-dominant SUSY breaking case with no anomalous $U(1)$ 
symmetry \cite{AU1}.
This nondegeneracy is dangerous for the FCNC.
We need the symmetries which constrain nonrenormalizable couplings, 
but do not lead to $D$-term contributions to soft scalar masses.

\vspace{0.3cm}

Superstring theory has such symmetries, i.e. stringy symmetries.
The orbifold construction is one of the simplest and most interesting 
constructions to derive four-dimensional string vacua \cite{Orbi}. 
In orbifold models, string states consist of the bosonic 
string on the four-dimensional space-time and a six-dimensional orbifold, 
their right-moving superpartners and left-moving gauge parts. 
This six-dimensional part and its supersymmetric part lead to 
complicated selection rules for allowed couplings \cite{Yukawa,Yukawa2,NR}.
In orbifold models each matter field corresponds to a 
$\theta^k$-twisted sector $T_k$ as well as the untwisted sector.
Stringy symmetries constrain couplings among these sectors.
Allowed nonrenormalizable couplings have been shown explicitly 
in Ref. \cite{NR2}.
Further orbifolds have some singular points, i.e., fixed points for 
each $T_k$ sector.
Each matter field is assigned to one of fixed points in $T_k$.
Stringy symmetries constrain combinations of fixed points 
for nonvanishing couplings \cite{Yukawa2}.
These selection rules for nonrenormalizable couplings cannot be 
understood in terms of effective field theories \cite{NR,NR2}.
Thus, if we assign the matter fields to these sectors and fixed points 
in certain way, we can obtain realistic fermion mass matrices.

\vspace{0.3cm}

Actually in Ref. \cite{stringm} an instructive pattern of quark mass 
matrices has been obtained as 
\begin{equation}
M_{\rm u,d} \; = \; c_{\rm u,d}
\pmatrix{
\varepsilon_{11} & \varepsilon_{\rm u,d}^3 & \varepsilon_{13} \cr
\varepsilon_{\rm u,d}^3 & \varepsilon_{\rm u,d}^2 
& \varepsilon_{\rm u,d}^2 \cr
\varepsilon_{13} & \varepsilon_{\rm u,d}^2 & 1 \cr
} \; ,
\end{equation}
up to $h_{{\rm u}ij}/h_{{\rm u}33}$ ($h_{{\rm d}ij}/h_{{\rm d}33}$) 
of $O(1)$, where 
$c_{\rm u}=h_{{\rm u}33}\langle H_2 \rangle$ and 
$c_{\rm d}=h_{{\rm d}33}\langle H_1 \rangle$.
For instance, this form of mass matrix is obtainable in the case 
where the Higgs fields are assigned to $T_4$ and the first, second 
and third families of quarks are assigned to $T_1$, $T_2$ and $T_4$ 
respectively, provided certain fields in $T_1$, $T_2$ and $T_4$ 
develop VEVs.
In this treatment, the (1,1), (1,3) and (3,1) matrix elements do not 
vanish completely. However, these elements are strongly suppressed 
in comparison with their nearest-neighboring elements, thus they
can always be taken as zero for any phenomenological purpose.

\vspace{0.3cm}

In general, we expect $\varepsilon_{\rm u} \neq \varepsilon_{\rm d}$. 
For example, the mixing between light 
and heavy Higgs fields leads to 
$\varepsilon_{\rm u} \neq \varepsilon_{\rm d}$ \cite{RRR,IG}.
The coupling strengths $h_{{\rm u}ij}$ ($h_{{\rm d}ij}$) are calculated 
as $h_{{\rm u}ij}\sim \exp(-a_{ij}T)$ 
($h_{{\rm d}ij}\sim \exp(-a'_{ij}T)$), where $T$ is the moduli 
parameter representing the size of six-dimensional compactified space 
and $a_{ij}$ ($a'_{ij}$) is a constant depending on the
combination of fixed points for couplings \cite{Yukawa} 
\footnote{Similarly the $CP$-violating phases can be introduced 
into some elements in the case of nonvanishing background antisymmetric 
tensors \cite{anti}, although without such antisymmetric tensors 
$CP$ is a nice symmetry of superstring and couplings are always real 
except trivial phases \cite{CP}.}.
The factors $\exp(-a_{ij}T)$ in the mass matrix elements are generally 
different from one another. Thus we phenomenologically introduce two 
additional parameters $\omega_{\rm u,d}$ for the (2,2) elements and 
discuss mass matrices of the form
\begin{equation}
M_{\rm u,d} \; = \; c_{\rm u,d}
\pmatrix{
0 & \varepsilon_{\rm u,d}^3 & 0 \cr
\varepsilon_{\rm u,d}^3 & \omega_{\rm u,d} ~ \varepsilon_{\rm u,d}^2 
& \varepsilon_{\rm u,d}^2 \cr
0 & \varepsilon_{\rm u,d}^2 & 1 \cr
} \; .
\end{equation}
Although $\omega_{\rm u,d} \sim O(1)$ are taken as a phenomenological 
assumption, we hope that they will have a theoretical explanation, 
analogous to that for $\varepsilon_{\rm u,d}$.   
In fact, we can assign fixed points for the matter fields so that 
such factors appear only for the (2,2) matrix elements.
Of course an introduction of similar factors for the other matrix
elements of $M_{\rm u,d}$ is not interesting in phenomenology.

\vspace{0.3cm}

Clearly the pattern of $M_{\rm u,d}$ can be regarded as a 
non-trivial generalization of the Fritzsch {\it Ansatz} 
\cite{Fr,FrXing} at the string scale $M_{\rm st}=3.7 \times 
10^{17}$ GeV. Diagonalizing $M_{\rm u,d}$ through the
orthogonal transformations $O^{\rm T}_{\rm u,d} M_{\rm u,d} O_{\rm u,d}
= {\rm Diag}\{ m_{u,d}, ~ m_{c,s}, ~ m_{t,b} \}$,
we are able to obtain the mass eigenvalues. 
In lowest order approximations, we find
\begin{eqnarray}
m_{t} & \approx & c_{\rm u} \; , ~~~~~~~~~~
m_{c} \; \approx \; \varepsilon_{\rm u}^{2} ~ \omega_{\rm u} ~ c_{\rm u} \; , 
~~~~~~~~~~
m_{u} \; \approx \; \frac{\varepsilon_{\rm u}^4}{\omega_{\rm u}} ~ c_{\rm u} 
\; ; \nonumber \\
m_{b} & \approx & c_{\rm d} \; , ~~~~~~~~~~
m_{s} \; \approx \; \varepsilon_{\rm d}^{2} ~ \omega_{\rm d} ~ c_{\rm d} \; , 
~~~~~~~~~~
m_{d} \; \approx \; 
\frac{\varepsilon_{\rm d}^4}{\omega_{\rm d}} ~ c_{\rm d} \; .
\end{eqnarray}
These lead to the following quark mass relations:
\begin{equation}
\frac{\omega_{\rm u}^2 ~ m_{u}}{m_{c}}\; \approx \; 
\frac{m_{c}}{\omega_{\rm u} ~  m_{t}} \; 
\approx \; \varepsilon_{\rm u}^{2} \; ,
~~~~~~~~~~
\frac{\omega_{\rm d}^2 ~ m_{d}}{m_{s}}\; \approx \; 
\frac{m_{s}}{\omega_{\rm d} ~ m_{b}} \; 
\approx \; \varepsilon_{\rm d}^{2} \; .
\end{equation}
Although simpler and more instructive geometrical relations for $m_u$,
$m_c$, $m_t$ and $m_d$, $m_s$, $m_b$ can be respectively obtained if one takes 
$\omega_{\rm u}=\omega_{\rm d}=1$, the latter will not be favored by current 
data at (or below) the weak scale $M_Z =91.187$ GeV. This is a
phenomenological reason for the essential presence of free parameters
$\omega_{\rm u,d}$ in $M_{\rm u,d}$. 

\vspace{0.3cm}

To calculate the mixing matrix of quark flavors, 
we need introduce a phase matrix
$P = {\rm Diag}\left \{ 1, ~ e^{{\rm i} \phi}, ~ e^{{\rm i} \phi} \right \}$,
where $\phi$ denotes the possible phase difference between $M_{\rm u}$
and $M_{\rm d}$. 
Such a $CP$-violating phase may arise from the 
dynamical details of our fermion mass generation mechanism, 
e.g., the background antisymmetric tensors in orbifold models or 
imaginary VEVs of $\theta$. 
Phenomenologically
the existence of $\phi$ is necessary for the {\it Ansatz} to properly 
reproduce the Cabibbo angle and $CP$ violation. The flavor mixing matrix,
defined as $V \equiv O^{\rm T}_{\rm u} P O_{\rm d}$, takes the following
form in leading order approximations:
\begin{equation}
V \; \approx \; \left ( \matrix{
\displaystyle 1-\frac{1}{2}\varepsilon^{{\prime}^2}_{\rm d}   ~ 
& ~ \displaystyle \varepsilon^{\prime}_{\rm u} e^{{\rm i} \phi} - 
\varepsilon^{\prime}_{\rm d}  
& ~ \displaystyle \varepsilon^{\prime}_{\rm u} \left (\varepsilon^{2}_{\rm d}
- \varepsilon^2_{\rm u} \right ) e^{{\rm i}\phi} \cr\cr
\varepsilon^{\prime}_{\rm d} e^{{\rm i}\phi} - \varepsilon^{\prime}_{\rm u}
& \displaystyle \left ( 1-\frac{1}{2}\varepsilon^{{\prime}^2}_{\rm d} \right ) 
e^{{\rm i}\phi} 
& \displaystyle \left (\varepsilon^2_{\rm d} - \varepsilon^2_{\rm u} \right ) 
e^{{\rm i}\phi} \cr\cr
\varepsilon^{\prime}_{\rm d} \left ( \varepsilon^2_{\rm u} 
- \varepsilon^{2}_{\rm d} \right ) e^{{\rm i}\phi} 
& \displaystyle \left ( \varepsilon^2_{\rm u} - \varepsilon^2_{\rm d} \right ) 
e^{{\rm i}\phi}
& \displaystyle e^{{\rm i}\phi} 
} \right ) \; ,
\end{equation}
where $\varepsilon^{\prime}_{\rm u,d}=\varepsilon_{\rm u,d}/\omega_{\rm u,d}$. 
This result will be confronted with current experimental data at the weak
scale $M_Z$ in the next section.

\section{Quark mixing and $CP$ violation at the weak scale}

We are in a position to compare the flavor mixing matrix $V$ 
with low-energy data, so as to phenomenologically justify
the string-inspired quark mass matrices $M_{\rm u,d}$.
For this purpose, one has to run the results obtained at the string scale 
$M_{\rm st}$ to the weak scale $M_Z$. 
We assume the MSSM for spontaneous symmetry breaking below 
$M_{\rm st}$, and then make use of the corresponding renormalization-group 
equations for quark mass matrices and $V$. 

\vspace{0.3cm}

The one-loop renormalization group equations for quark
mass ratios and flavor mixing matrix elements have been explicitly 
presented by 
Babu and Shafi in Ref. \cite{BS}.
In view of the hierarchy of Yukawa couplings and quark mixing angles, one can 
make reliable analytical approximations for the relevant evolution equations 
by keeping only the leading terms. It has been found that 
(1) the running effects of $m_u/m_c$ and $m_d/m_s$ are negligibly small;
(2) the diagonal elements of the flavor mixing matrix have 
negligible evolutions 
with energy;
(3) the evolutions of $|V_{us}|$ and $|V_{cd}|$ involve 
the second-family Yukawa couplings and thus they are negligible; 
(4) the flavor mixing matrix elements $|V_{ub}|$, $|V_{cb}|$, $|V_{td}|$ 
and $|V_{ts}|$ have identical running behaviors. 
Taking these points into account, we first notice that 
$\varepsilon^{\prime}_{\rm u} \approx (m_u/m_c)^{1/2}$ and
$\varepsilon^{\prime}_{\rm d} \approx (m_d/m_s)^{1/2}$ are
approximately scale-independent. Consequently some scale-independent 
results for quark mixings at $M_Z$ can be straightforwardly obtained 
as follows.

\vspace{0.3cm}

(a) The magnitudes of three diagonal elements of $V$ are close to unity, i.e.,
$|V_{ud}| \approx |V_{cs}| \approx |V_{tb}| \approx 1$. 
A more careful estimate
with the help of unitarity of $V$ leads to the fine hierarchy 
$|V_{tb}| > |V_{ud}| > |V_{cs}|$. This result is consistent with  current 
data \cite{Xing}.

\vspace{0.3cm}

(b) The flavor mixing matrix elements $|V_{us}|$ and $|V_{cd}|$ 
in the leading order approximation read
\begin{equation}
|V_{us}| \; \approx \; |V_{cd}| \; \approx \; \left [ \frac{m_u}{m_c} 
+\frac{m_d}{m_s} - 2 \left ( \frac{m_u m_d}{m_c m_s} \right )^{1/2} 
~ \cos \phi \right ]^{1/2} \; .
\end{equation}
Since $|V_{us}|$ ($=0.2205 \pm 0.0018$ \cite{PDG}) has been accurately 
measured 
and a reliable value for $m_s/m_d$ ($=18.9\pm 0.8$ \cite{Leutwyler}) 
has been obtained from chiral perturbation theory, we are able to constrain 
the phase shift $\phi$ in spite of the large uncertainty associated with 
$m_u/m_c$ ($\sim 5 \times 10^{-3}$ \cite{PDG}). We find $73^0 \leq \phi
\leq 82^0$ only if $m_u/m_c \geq 10^{-3}$. 

\vspace{0.3cm}

(c) The ratios $|V_{ub}/V_{cb}|$ and $|V_{td}/V_{ts}|$ are obtained as
\begin{equation}
\left | \frac{V_{ub}}{V_{cb}} \right | \; \approx \; \left ( \frac{m_u}{m_c}
\right )^{1/2} \; , ~~~~~~~~
\left | \frac{V_{td}}{V_{ts}} \right | \; \approx \; \left ( \frac{m_d}{m_s}
\right )^{1/2} \; ,
\end{equation}
to a good degree of accuracy. By use of $m_s/m_d = 18.9\pm 0.8$ 
\cite{Leutwyler}, we get $0.225\leq |V_{td}/V_{ts}| \leq 0.235$. 
This result is consistent with that extracted from the analysis
of current experimental data \cite{AL}: $0.15\leq |V_{td}/V_{ts}|\leq 0.34$.
On the other hand, the allowed region of $|V_{ub}/V_{cb}|$ is restricted 
by that of $m_u/m_c$, which has not been reliably determined. 
We find that $0.0036 \leq m_u/m_c \leq 0.01$ is required by our 
{\it Ansatz} in fitting the data $|V_{ub}/V_{cb}| = 0.08\pm 0.02$ \cite{PDG}. 

\vspace{0.3cm}

(d) The unitarity triangle $V^*_{ub}V_{ud} + V^*_{cb}V_{cd} 
+ V^*_{tb}V_{td} =0$ 
can be approximately derived from our quark mass {\it Ansatz}. From Eqs. (6) 
and (7) we observe that 
$V_{cd}$, $(m_u/m_c)^{1/2}$ and $(m_d/m_s)^{1/2}$ form a triangle
in the complex plane \cite{FrXing}. 
Rescaling three sides of this triangle
by $V^*_{cb}$ and making use of Eq. (8), we then reproduce the
above-mentioned unitarity triangle in leading order approximations.
Thus three inner angles of the unitarity triangle, commonly denoted as
$\alpha$, $\beta$ and $\gamma$ \cite{PDG}, can be calculated in terms of
quark mass ratios and $\phi$. 
In $B$-meson physics, one is more interested
in the characteristic $CP$-violating observables $\sin(2\alpha)$, 
$\sin(2\beta)$ 
and $\sin\gamma$, which will be detected at the forthcoming $B$ factories.
Explicitly, we find
\begin{eqnarray}
\sin(2\alpha) & \approx & \sin(2\phi) \; , \nonumber \\
\sin(2\beta) & \approx & \frac{2\sin\phi ~ (r - \cos\phi)}
{1 - 2r \cos\phi + r^2} \; , \nonumber \\
\sin^2 \gamma & \approx & \frac{r^2 \sin^2\phi}{1 - 2r \cos\phi + r^2} \; , 
\end{eqnarray}
where $r = [(m_c m_d)/(m_u m_s)]^{1/2}$. A brief estimate gives
$0.18\leq \sin(2\alpha) \leq 0.58$, $0.5\leq \sin(2\beta) \leq 0.78$
and $0.93\leq \sin^2\gamma \leq 1.0$, which are very restrictive but
consistent with current data analyzed by Ali and London \cite{AL}:
$-0.90\leq \sin(2\alpha) \leq 1.0$, $0.32\leq \sin(2\beta) \leq 0.94$
and $0.34 \leq \sin^2\gamma \leq 1.0$. 

\vspace{0.3cm}

We want to stress that all the scale-independent results obtained above 
are indeed the consequences of the Fritzsch {\it Ansatz} \cite{Fr}.
They appear naturally only if the (1,1), (1,3) and (3,1) elements of
$M_{\rm u,d}$ are greatly suppressed in comparison with their 
nearest-neighboring elements \cite{Hall,SandaXing}. Next we discuss the
scale dependence of $|V_{cb}|$ and $|V_{ts}|$, whose magnitudes in
our {\it Ansatz} are absolutely different from those in the Fritzsch
{\it Ansatz} or other quark mass patterns \cite{Xing96,96mass}. 

\vspace{0.3cm}

Non-negligible running effects from the string scale $M_{\rm st}$ to 
the weak scale $M_Z$ can 
manifest themselves in the expressions of $|V_{cb}|$ and $|V_{ts}|$, 
which depend strongly upon the mass ratios $m_c/m_t$ and $m_s/m_b$. 
The evolution functions 
relevant to the calculation of $|V_{cb}|$ or $|V_{ts}|$ are defined as
\begin{equation}
\xi_{t,b} \; = \; \exp \left [ -\frac{1}{16\pi^2} 
\int^{\ln (M_{\rm st}/M_Z)} _0 f^2_{t,b}(\chi) ~ {\rm d}\chi \right ] \; ,
\end{equation}
where $\chi \equiv \ln (\mu /M_Z)$, $f_t$ and $f_b$ are the respective Yukawa
coupling eigenvalues of the top and bottom quarks.
A good approximation is that the third-family Yukawa couplings 
of quarks and charged leptons, together with the gauge couplings, 
play the dominant roles in the renormalization-group equations \cite{BS}.
Then the magnitudes of $\xi_t$ and $\xi_b$ can be evaluated for arbitrary
$\tan\beta_{\rm susy}$ from $M_{\rm st}$ to $M_Z$, as done in \cite{Xing96}
with the typical inputs $m_t (M_Z) \approx 180$ GeV, $m_b (M_Z) \approx
3.1$ GeV and $m_{\tau} (M_Z) \approx 1.78$ GeV. For our present purpose, 
the numerical results of $\xi_t$ and $\xi_b$ as functions of 
$\tan\beta_{\rm susy}$ are illustrated in Fig. 1.
Three key evolution relations in the MSSM are given as follows \cite{Xing96}:
\begin{eqnarray}
\left . \frac{m_s}{m_b} \right |_{M_Z} & = & \frac{1}{\xi_t ~ \xi^3_b} ~
\left . \frac{m_s}{m_b} \right |_{M_{\rm st}} \; , \nonumber \\
\left . \frac{m_c}{m_t} \right |_{M_Z} & = & \frac{1}{\xi^3_t ~ \xi_b} ~ 
\left . \frac{m_c}{m_t} \right |_{M_{\rm st}}  \; , \nonumber \\
\left |\hat{V}_{ij} \right |_{M_Z} & = & \frac{1}{\xi_t ~ \xi_b} ~ 
\left |\hat{V}_{ij} \right |_{M_{\rm st}} \; ,
\end{eqnarray}
where $(ij) = (ub)$, $(cb)$, $(td)$ or $(ts)$. With the help of 
Eq. (6), we obtain the renormalized $|V_{cb}|$ and
$|V_{ts}|$ at $M_Z$ in leading order approximations:
\begin{equation}
|V_{cb}| \; \approx \; |V_{ts}| \; \approx \;
\frac{\xi_b}{\xi_t^{1/3}} \left (\frac{m_d}{m_s} \right )^{1/3} 
\left (\frac{m_s}{m_b} \right )^{2/3} ~ - ~
\frac{\xi_t}{\xi_b^{1/3}} \left (\frac{m_u}{m_c} \right )^{1/3}
\left (\frac{m_c}{m_t} \right )^{2/3} \; .
\end{equation}
This instructive result is a unique consequence of our quark mass
{\it Ansatz} given in Eq. (3).
Typically taking $m_u/m_c=0.006$, $m_c/m_t=0.005$, 
$m_d/m_s = 0.051 \sim 0.055$ and $m_s/m_b=0.03 \sim 0.04$ 
\cite{Leutwyler,GL},
we confront Eq. (12) with the experimental data on $V_{cb}$
(i.e., $|V_{cb}|=0.0388\pm 0.0032$ \cite{Neubert}). 
As shown in Fig. 2, our result is in good agreement with experiments 
for $\tan\beta_{\rm susy} < 55$. This implies that
the quark mass matrices $M_{\rm u,d}$, proposed at the string
scale $M_{\rm st}$, may have a large chance to 
survive for reasonable values of $\tan\beta_{\rm susy}$.

\section{Summary}

We have derived a simple and realistic pattern of quark mass matrices 
from orbifold models of superstring theory at the string scale 
$M_{\rm st}$. It is worth emphasizing that gauge symmetries 
are not needed for our purpose. This point is so important
that one can preserve the degeneracy of squark masses without fine-tuning.
The obtained up and down mass matrices have the parallel structure, and
each of them are left-right symmetric. They totally consist of seven
free parameters, thus can lead to three independent predictions for
flavor mixing angles at the string scale. From the purely phenomenological
point of view, our quark mass pattern can be regarded as a non-trivial 
generalization of the Fritzsch {\it Ansatz}.

\vspace{0.3cm}

The consequences of our string-inspired mass matrices $M_{\rm u,d}$ 
on flavor mixing and $CP$ violation have been confronted with current
experimental data at the weak scale $M_Z$ by use of the one-loop 
renormalization-group equations in the MSSM framework. In leading order
approximations, we find that $|V_{us}|$, $|V_{cd}|$, $|V_{ub}/V_{cb}|$
and $|V_{td}/V_{ts}|$ are scale-independent. Consequently three inner
angles of the unitarity triangle $V^*_{ub}V_{ud} + V^*_{cb}V_{cd}
+ V^*_{tb}V_{td}=0$ are approximately scale-independent, and their 
magnitudes are independent of $m_c/m_t$ and $m_b/m_t$ to a good degree
of accuracy. We get very restrictive results 
$0.18 \leq \sin(2\alpha) \leq 0.58$, $0.5 \leq \sin(2\beta) \leq 0.78$
and $0.93 \leq \sin^2\gamma \leq 1.0$ (see also Ref. \cite{Xing96}),
which can be tested in the forthcoming experiments of $B$-meson
factories. 

\vspace{0.3cm}

The flavor mixing matrix elements, which are most sensitive to the
features of different
quark mass $Ans\ddot{a}tze$, are usually $|V_{cb}|$ and $|V_{ts}|$. 
Our mass pattern yields a unique expression for $|V_{cb}|$ (or $|V_{ts}|$)
in terms of quark mass ratios $m_u/m_c$, $m_c/m_t$, $m_d/m_s$ and
$m_s/m_b$. At the weak scale $M_Z$, the renormalized $|V_{cb}|$ can
fit current data very well for reasonable values of $\tan\beta_{\rm susy}$.
Of course, all numerical results depend upon the inputs of quark mass
eigenvalues, which still have large uncertainties. 

\vspace{0.3cm}

The good agreement between the string-inspired mass pattern and current 
data implies that the former may have a large chance to be true. To
reduce the number of free parameters in our {\it Ansatz}, a possible
way is to relate the quark mass matrices with the lepton mass matrices
at the string scale. Then one should be able to predict the quark 
masses in terms of the lepton masses with the help of some discrete
symmetries (see, e.g., Ref. \cite{GJ}). Such possibilities are of course
attractive from both theoretical and phenomenological viewpoints, and they
will be studied elsewhere.

\vspace{0.5cm}

\begin{flushleft}
{\Large\bf Acknowledgements}
\end{flushleft}

One of the authors (ZZX) would like to thank A.I. Sanda for his
warm hospitality and the Japan Society for the Promotion of
Science for its financial support. 
This work was started when both of the authors visited the Physics 
Department of Universit$\rm\ddot{a}$t M$\rm\ddot{u}$nchen as the 
Alexander von Humboldt research fellows.

\vspace{0.8cm}

\end{document}